\documentclass[journal,letterpaper,twocolumn]{IEEEtran}
\usepackage{amsfonts}
\usepackage{amssymb}
\usepackage{cite}
\usepackage[cmex10]{amsmath}
\usepackage{float}
\usepackage{color}
\usepackage{stfloats,fancyhdr}

\usepackage{algorithm}
\usepackage{algorithmic}
\usepackage{multirow}
\usepackage{changepage}
\usepackage[normalem]{ulem}

\newtheorem{proposition}{Proposition}
\newtheorem{corollary}{Corollary}

\newtheorem{remark}{Remark}

\IEEEoverridecommandlockouts

\ifCLASSINFOpdf
  \usepackage[pdftex]{graphicx}
  \DeclareGraphicsExtensions{.pdf,.jpeg,.png}
\else
  \usepackage[dvips]{graphicx}
  \DeclareGraphicsExtensions{.eps}
\fi

\usepackage{subfigure}

\usepackage{fancybox,dashbox}

\begin{document}

\title{Harvest-Then-Cooperate: Wireless-Powered Cooperative Communications
\thanks{Copyright (c) 2015 IEEE. Personal use of this material is permitted. However, permission to use this material for any other purposes must be obtained from the IEEE by sending a request to pubs-permissions@ieee.org.}
\thanks{The work was supported by the Australian Research Council (ARC) under Grants DP120100190 and FT120100487, International Postgraduate Research Scholarship (IPRS), Australian Postgraduate Award (APA), and Norman I Price Supplementary Scholarship. Part of this work was presented at the IEEE Global Communications Conference (Globecom), Austin, TX USA, 8-12 December, 2014 \cite{Chen_Globecom_2014_A}.}
\thanks{H. Chen, Y. Li, and B. Vucetic are with School of Electrical and Information Engineering, The University of Sydney, Sydney, NSW 2006, Australia (email: he.chen@sydney.edu.au, yonghui.li@sydney.edu.au, branka.vucetic@sydney.edu.au).}
\thanks{J. L. Rebelatto is with Federal University of Technology-Parana, Curitiba, PR, 80230-901, Brazil (email: jlrebelatto@utfpr.edu.br).
}
\thanks{
B. F. Uch\^{o}a-Filho is with Federal University of Santa
Catarina, Florian\'{o}polis, SC, 88040-900, Brazil (email: uchoa@eel.ufsc.br).}
}
\author{He~(Henry)~Chen,~\IEEEmembership{Student Member,~IEEE,}
Yonghui Li,~\IEEEmembership{Senior Member,~IEEE,} \\
Jo\~{a}o Luiz Rebelatto,~\IEEEmembership{Member,~IEEE,}~Bartolomeu F. Uch\^{o}a-Filho,~\IEEEmembership{Senior Member,~IEEE,}\\Branka Vucetic,~\IEEEmembership{Fellow,~IEEE}
}

\maketitle

\begin{abstract}
In this paper, we consider a wireless-powered cooperative communication network consisting of one hybrid access-point (AP), one source, and one relay. In contrast to conventional cooperative networks, the source and relay in the considered network have no embedded energy supply. They need to rely on the energy harvested from the signals broadcasted by the AP for their cooperative information transmission. Based on this three-node reference model, we propose a harvest-then-cooperate (HTC) protocol, in which the source and relay harvest energy from the AP in the downlink and work cooperatively in the uplink for the source's information transmission. Considering a delay-limited transmission mode, the approximate closed-form expression for the average throughput of the proposed protocol is derived over Rayleigh fading channels. Subsequently, this analysis is extended to the multi-relay scenario, where the approximate throughput of the HTC protocol with two popular relay selection schemes is derived. The asymptotic analyses for the throughput performance of the considered schemes at high signal-to-noise radio are also provided. All theoretical results are validated by numerical simulations. The impacts of the system parameters, such as time allocation, relay number, and relay position, on the throughput performance are extensively investigated.
\end{abstract}


\IEEEpeerreviewmaketitle

\section{Introduction}

As a sustainable solution to prolonging the lifetime of energy constrained wireless networks, the energy harvesting technique has recently drawn significant attention, e.g. see \cite{Ozel_JSAC_2011} and references therein. It enables wireless nodes to collect energy from the surrounding environment. Apart from the conventional renewable energy sources such as solar and wind, radio frequency (RF) signals radiated by ambient transmitters can be treated as a viable new source for energy harvesting. Thus, the wireless signals can be used to deliver information as well as energy. In recent years, some significant advances in wireless power technologies have highly increased the feasibility of wireless energy transfer in practical wireless applications \cite{Shinohara_mag_2011_Power}. As an example, the successful communication between two terminals solely powered by ambient radio signals, such as the existing TV and cellular signals, has been realized and reported in \cite{Liu_SIGCOMM_2013_Ambient}. Besides ambient radio signals, dedicated power transmitters are deployed to implement wireless energy transfer in some applications, e.g., passive radio frequency identification (RFID) networks \cite{Smith_book_2013_wirelessly}. Moreover, with further advances in antenna technology and energy harvesting circuit designs, wireless energy transfer is believed to be more efficient such that it will be implemented widely in the near future.

In this context, a new type of wireless network, termed wireless-powered communication network (WPCN), has become a promising research topic and attracted more and more attention. In WPCNs, the wireless terminals are powered only by wireless energy transfer and transmit their information using the harvested energy. The WPCNs under different setups have been studied in open literature. Specifically, \cite{Huang_TWC_2014_Enabling} proposed a new network architecture to enable wireless energy transfer in hybrid cellular networks, where an uplink cellular network overlays with randomly deployed power beacons for powering mobiles by microwave radiation. In this paper, the deployment of this hybrid network under an outage constraint on data links was designed using the stochastic-geometry theory. In \cite{Nintanavongsa_INFOCOM_13_Medium}, a medium access control (MAC) protocol was proposed for sensor networks powered by wireless energy transfer. Wireless energy transfer was considered for cognitive radio networks in \cite{Lee_TSP_2013_opp}, where secondary transmitters harvest ambient RF energy from transmissions by nearby active primary transmitters.

A classic multi-user WPCN was first investigated in \cite{Ju_TWC_2014}. In this work, a ``harvest-then-transmit" protocol was developed, where the users first collect energy from the signals broadcasted by a single-antenna hybrid access-point (AP) in the downlink (DL) and then use their harvested energy to send \emph{independent} information to the hybrid AP in the uplink (UL) based on the TDMA. Very recently, the full-duplex technique was adopted to further improve the performance of a multi-user WPCN in \cite{Ju_arXiv_2014_Optimal, Kang_arXiv_2014_Full}, where the hybrid AP implements the full-duplex through two antennas: one for broadcasting wireless energy to users in the DL and the other for receiving independent information from users via TDMA in the UL at the same time. Moreover, the throughput of a massive MIMO WPCN was optimized in \cite{Yang_arXiv_2014_Thro}, in which the hybrid AP is assumed to be equipped with a large number of antennas.

There are also significant amount of research papers that focused on the design of the simultaneous wireless information and power transfer (SWIPT) (e.g., \cite{Varshney_ISIT_2008,Zhou_J_2012,Zhou_TWC_2013,Ng_TWC_2013_Wireless,Huang_TSP_2013_Simu,Ding_TWC_2014_Power,Krikidis_Tcom_2013,Chen_ISIT_2014_A,Chen_TWC_2015_Distributed} and references therein). These works considered that wireless energy transfer and wireless information transmission are in the same direction and focused on the characterization of the fundamental tradeoffs between energy transfer and information transmission.

On the other hand, cooperative diversity technique has rekindled enormous interests from the wireless communication community over the past decade. The key idea of this technique is that single-antenna nodes in wireless networks share their antennas and
transmit cooperatively as a virtual multi-input multi-output (MIMO) system, thus spatial diversity can be achieved without the need of multiple antennas at each node \cite{Laneman_TIT_2004,Yonghui_book_2010}. The advantages of this technique, such as increasing system capacity, coverage and energy efficiency, have been demonstrated by numerous papers in open literature.  Very recently, the concept of user cooperation was applied to a WPCN in \cite{Moritz_TSP_2013_sub}, where a (hybrid) destination node first charges two cooperative source nodes with wireless energy transfer and then collects information from them. A time-switching network-coded cooperative protocol was proposed, by which two source nodes can cooperatively transmit their information in two consecutive transmission blocks. The outage performance of the considered system was analyzed and optimized with respect to the time allocation parameter \cite{Moritz_TSP_2013_sub}. In cooperative networks, there is another (asymmetric) model using relay node(s) to assist information forwarding, which has many applications in practice. However, to the best of our knowledge, there is no paper that considered the design of the aforementioned model for WCPNs in open literature. This gap actually motivates this paper.

\begin{figure}
\centering \scalebox{0.55}{\includegraphics{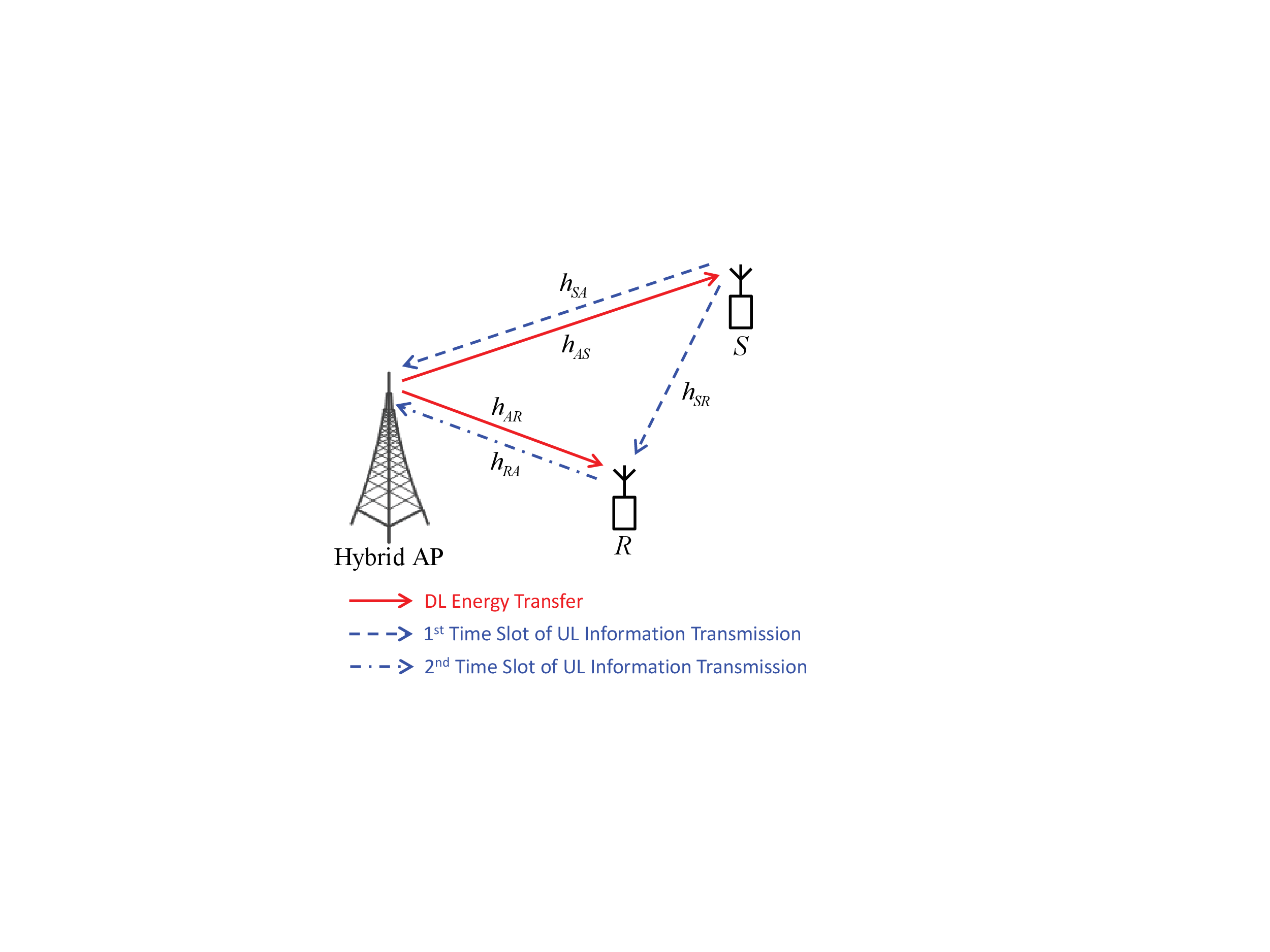}}
\caption{A reference model for wireless-powered cooperative communication network (WPCCN) with energy transfer in the DL and cooperative information transmission in the UL. \label{fig:system_model}}
\end{figure}
In this paper, we study a time-switching cooperative communication network with DL wireless energy transfer (WET) and UL wireless information transmission (WIT). As shown in Fig. \ref{fig:system_model}, the considered network consists of one hybrid AP, one (information) source ($S$), and another terminal named as relay ($R$) that has no traffic and is willing to assist the information transmission of the source. The hybrid AP is connected to a constant power supply, while the source and relay are assumed to have no other energy sources. But they are equipped with a rechargeable battery and thus can harvest and store the wireless energy broadcasted by the hybrid AP. Unlike prior work on the design of SWIPT in relay networks (e.g., \cite{Zhou_TWC_2013,Ding_TWC_2014_Power}) that focused on the WET and WIT in the same direction, we consider the scenario with WET in the DL and WIT in the UL\footnote{A more general setup is that with both WET and WIT in the DL and WIT in the UL. For the purpose of exposition, we ignore the DL WIT in this paper.}. Particularly, the hybrid AP broadcasts only wireless energy to the source and relay in the DL, while the source and relay cooperatively transmit the source's information using their individual harvested energy to the hybrid AP in the UL. We refer to the considered network as a wireless-powered cooperative communication network (WPCCN) in practice.

The main contributions of this paper are summarized as follows:
\begin{itemize}
  \item Based on the three-node reference model depicted in Fig.~\ref{fig:system_model}, we propose a harvest-then-cooperate (HTC) protocol for the WPCCN, where the source and relay harvest energy from the AP during the DL phase and cooperate for the source's information transmission during the UL phase. The amplify-and-forward (AF) relaying scheme \cite{Laneman_TIT_2004} and the selection combining technique \cite{Simon_book_2000} are assumed to be implemented at the relay and the AP, respectively. Considering the delay-limited transmission mode \cite{Zhou_TWC_2013}, we derive the approximate closed-form expression of the average throughput for the proposed HTC protocol over Rayleigh fading channels. %Since the signal-to-noise ratios (SNRs) of source-AP and source-relay-AP are correlated due to the WET process in the WPCCN, its performance analysis becomes a non-trivial task, requiring different analytical approach compared to the ones for conventional cooperative networks.
  \item We subsequently extend the analysis to the multi-relay scenario, where the single relay selection technique is implemented. In particular, we consider that only one of the relays will be selected according to a certain criterion in each transmission block, and the selected relay will use the energy harvested in this block to forward the received signal from the source. Two popular relay selection schemes, i.e., opportunistic relaying and partial relay selection, are considered and the corresponding throughput performances are also analyzed.
  \item All theoretical results are validated by numerical simulations. The impacts of the system parameters, such as time allocation, relay number, and relay position, on the throughput performance of the considered schemes are extensively investigated. Numerical results show that the proposed scheme outperforms the existing  harvest-then-transmit protocol \cite{Ju_TWC_2014} in all simulated cases.
\end{itemize}

It is worth emphasizing that the analytical approach adopted in this paper is technically different from the existing ones for conventional cooperative networks. One difference is that the signal-to-noise ratio (SNR) of each link in the considered WPCCN is proportional to the product of two exponential random variables instead of one variable as in conventional cooperative networks. However, this is not the principal technical difference: due to the inherent energy transfer process in the considered WPCCN, the source's transmit power in the uplink is a random variable instead of a constant as in conventional cooperative networks. This makes the SNRs of the source-AP link and all source-relay-AP links mutually correlated, which is essentially different from the conventional cooperative networks with independent link SNRs. As a result, the analytical tools for conventional cooperative networks cannot be directly used here.

The rest of this paper is organized as follows. The system model and the proposed HTC protocol are described in Section~II. Section III derives the approximate closed-form expression for the average throughput of the proposed protocol in the three-node reference model. The extension to the multi-relay scenario is discussed and analyzed in Section IV. The simulation results are presented in Section V to validate the theoretical analyses and demonstrate the impacts of the system parameters. Finally, Section VI concludes the paper.

\textbf{\emph{Notations:}} Throughout this paper, $\mathbb{E}\left\{ \cdot \right\}$ and $\left| {\cdot} \right|$ denote the expectation and the absolute value operations, respectively. $X \sim {\mathcal {CN}}\left( {\mu ,{\sigma ^2}} \right)$ stands for a circularly symmetric complex Gaussian random variable $X$ with mean $\mu$ and variance ${\sigma ^2}$, while $X \sim {\rm {EXP}}\left( {{\lambda}} \right)$ represents an exponentially distributed random variable $X$ with mean ${\lambda}$. $ \Pr \left( A \right)$ and $\Pr \left( {A,B} \right) $ denote the probability that the event $A$ happens and the probability that the events $A$ and $B$ happen simultaneously, respectively.

\section{System Model and Protocol Description}
As shown in Fig. \ref{fig:system_model}, this paper considers a WPCCN with energy transfer in the DL and cooperative information transmission in the UL. It is assumed that all nodes are equipped with one single antenna and work in the half-duplex mode. In addition, the source and relay are assumed to have no other embedded energy supply and thus need to first harvest energy from the signal broadcasted by the AP in the DL, which can be stored in a rechargeable battery and then used for the information transmission to the AP in the UL.

In the sequel, we use subscript-$A$ for AP, subscript-$S$ for source, and subscript-$R$ for relay. We use ${\tilde h_{XY}}\sim {\mathcal {CN}}\left( {0 ,{\sigma_{XY} ^2}} \right)$ to denote the channel coefficient from $X$ to $Y$ with $X,Y \in \left\{ {A,S,R} \right\}$. The channel power gain ${h_{XY}} = {\left| {{{\tilde h}_{XY}}} \right|^2}$ from $X$ to $Y$ thus follows the exponential distribution with the mean ${\sigma_{XY} ^2}$, i.e., ${h_{XY}}\sim {\rm {EXP}}\left( {{{\sigma_{XY} ^2}}} \right)$ \cite{Simon_book_2000}. In addition, it is assumed that all channels in both DL and UL experience independent slow and frequency flat fading, where the channel gains remain constant during each transmission block (denoted by $T$) but change independently from one block to another.

\begin{figure}
\centering \scalebox{0.32}{\includegraphics{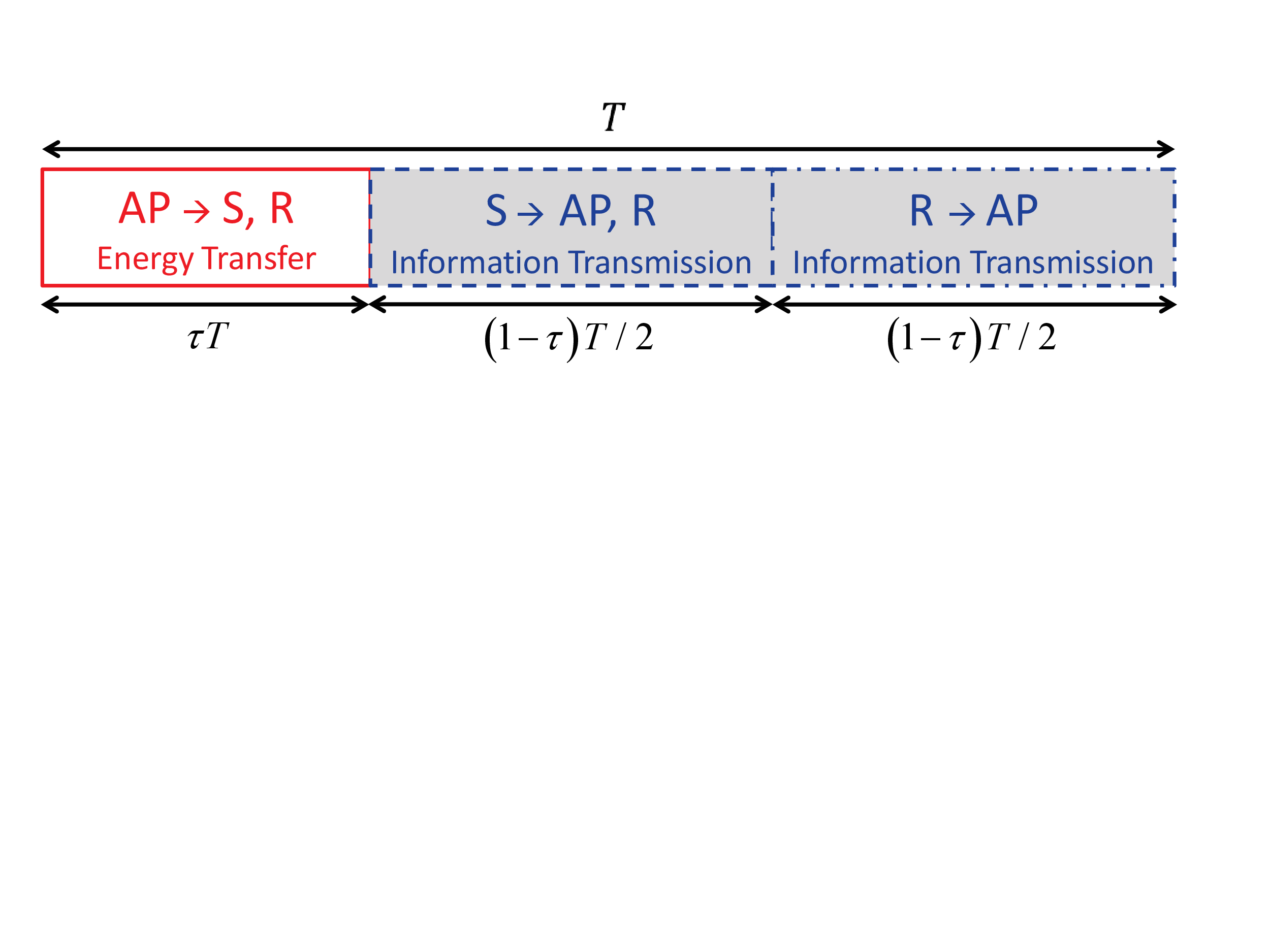}}
\caption{Diagram of the harvest-then-cooperate protocol. \label{fig:time_allocation}}
\end{figure}
The proposed \emph{harvest-then-cooperate} (HTC) protocol for the considered network is shown in Fig. \ref{fig:time_allocation}. Specifically, in each transmission block of time duration $T$, the first $\tau T$ amount of time with $0<\tau<1$ is assigned to the DL energy transfer from the AP to the source and relay. The remaining fraction $1-\tau$ of the block is further divided into two time slots with the equal length of $\left(1-\tau\right)T/2$ for cooperative information transmission in the UL. During the first time slot of the UL, the source uses the harvested energy to transmit data information to the AP, which can also be overheard by the relay due to the broadcasting feature of wireless communication. In the second time slot of the UL, the relay will use the energy harvested during the DL phase to help forwarding the source information through employing the
amplify-and-forward (AF) relaying protocol due to its lower complexity\footnote{In this paper, the possibility of the source harvesting energy during the relay's transmission is not taken into account. Such amount of energy is neglected since, as commented in \cite{Ishibashi_PIMRC_2012_Energy,Moritz_TSP_2013_sub}, the energy transfer efficiency is maximized for narrowband links that operate at low frequencies. However, the relay's data transmission needs a high data rate with even larger bandwidth and consequently leads to a lower energy harvesting efficiency.} \cite{Laneman_TIT_2004}. 
To reduce the complexity of the receiver structure, we assume that the selection combining (SC) technique \cite{Simon_book_2000} is implemented for information processing at the AP. Moreover, the SC technique enables the tractable closed-form analysis of the considered system. Specifically, the information receiver of the AP will select one of the received signals from the source and relay during the UL phase, which has larger SNR, to decode the source's information at the end of each block.

Let $P_A$ denote the transmission power of the AP during the DL phase. Also, we assume that $P_A$ is sufficiently large such that the energy harvested from the noise is negligible. Thus, the amount of energy harvested by the source and relay can be expressed as \cite{Ju_TWC_2014}
\begin{equation}\label{eq:harvested_energy_source}
{E_S} = \eta \tau T {P_A}h_{AS},
{~\rm{and}}~
{E_R} = \eta \tau T {P_A}h_{AR},
\end{equation}
where $0<\eta<1$ is the energy harvesting efficiency. For convenience but without loss of generality, we consider a normalized unit block time (i.e., $T=1$) hereafter.

After the terminals replenish their energy during the DL phase, they work cooperatively for the information transmission in the subsequent UL phase. For the purpose of exploration, we follow \cite{Ju_TWC_2014,Zhou_TWC_2013} and assume that both terminals exhaust the harvested energy for uplink information transmission. Note that instead of exhausting their harvested energy, the source and relay can also choose to perform power control (allocation) across different transmission blocks to further improve the network performance, which, however, is out the scope of this paper. On the other hand, the network throughput performance derived in this paper can actually be treated as the lower bound of the aforementioned power control scenario.

The transmit power of the source and relay during the UL phase are thus given by
\begin{equation}\label{}
{P_S} = {{{E_S}}}/ \left[{{\left( {1 - \tau } \right)/2}}\right] = {{2\eta \tau {P_A}h_{AS}}}/\left({{1 - \tau }}\right),
\end{equation}
\begin{equation}\label{eq:relay_transmit_power}
{P_R} = {{{E_R}}}/ \left[{{\left( {1 - \tau } \right)/2}}\right] = {{2\eta \tau {P_A}h_{AR}}}/\left({{1 - \tau }}\right).
\end{equation}
Thus, the received SNR at the AP after the source's transmission can be written as
\begin{equation}\label{eq:SNR_H_source}
{\gamma _{SA}} = {{{P_S}{{ {{h_{SA}}} }}}}/{{{N_0}}} = \mu {{ {{h_{AS}}} }}{{ {{h_{SA}}} }},
\end{equation}
where $N_0$ is the power of the noise suffered by all receivers and
\begin{equation}\label{eq:def_mu}
\mu = {{2\eta  \left({P_A}/{N_0}\right)\tau}}/{{\left( {1 - \tau } \right)}}.
\end{equation}

At the same time, the signal sent by the source can also be overheard by the relay. In the second time slot of the UL phase, the relay will amplify and forward the received signal to the AP using the power $P_R$ given in (\ref{eq:relay_transmit_power}) and the amplification factor $\beta  = 1/\sqrt {{P_S}{h_{SR}} + {N_0}} $ \cite{Ikki_CL_2007}. After some algebraic manipulations, we can express the received SNR at the AP from the link $S$-$R$-$A$ as
\begin{equation}\label{eq:exact_gamma_SRA}
{\gamma _{SRA}} = \frac{{\mu {h_{AS}}{h_{SR}}\mu{h_{AR}}{h_{RA}}}}{{\mu{h_{AS}}{h_{SR}} +\mu {h_{AR}}{h_{RA}} + 1 }}.
\end{equation}

Since the SC technique is adopted at the AP receiver, the output SNR of the HTC protocol is given by
\begin{equation}\label{}
{\gamma _A } = \max \left( {{\gamma _{SA}},{\gamma _{SRA}}} \right).
\end{equation}

In the following section, we will analyze the average throughput performance of the proposed HTC protocol in the three-node reference model depicted in Fig. \ref{fig:system_model}.

\section{Throughput Analysis of the HTC Protocol}\label{sec:analysis}
In this paper, we consider the delay-limited transmission mode, where the average throughput can be obtained by evaluating the outage probability of the system with a fixed transmission rate \cite{Zhou_TWC_2013}. Specifically, the throughput of a given system with transmission rate $R$, outage probability $P_{\rm{out}}$ and transmission duration $t$ is given by $R(1-P_{\rm{out}})t$.

To analyze the throughput of the HTC protocol in the three-node reference model, we first characterize the outage probability of the proposed protocol and have the following proposition:
\begin{proposition}\label{prop:outage_single_relay}
The outage probability of the HTC protocol can be approximately expressed as
\begin{equation}\label{eq:outage_HTC_AF_prop}
\begin{split}
&P_{{\rm{out}}}^{\rm{HTC}} \approx 1 - {\cal S}\left( {\frac{{4\nu }}{{\mu \sigma _{AS}^2\sigma _{SA}^2}}} \right) - {\cal S}\left( {\frac{{4\nu }}{{\mu \sigma _{AR}^2\sigma _{RA}^2}}} \right) \times  \\
&\left[ {{\cal S}\left( {\frac{{4\nu }}{{\mu \sigma _{AS}^2\sigma _{SR}^2}}} \right) - {\cal S}\left( {\frac{{4\nu }}{{\mu \sigma _{AS}^2\sigma _{SR}^2}} + \frac{{4\nu }}{{\mu \sigma _{AS}^2\sigma _{SA}^2}}} \right)} \right],
\end{split}
\end{equation}
where
\begin{equation}\label{eq:S_func}
{\cal S}\left( x \right) = \sqrt x {K_1}\left( {\sqrt x } \right)
\end{equation}
is defined for notation simplification with $K_1\left(\cdot\right)$ denoting the modified Bessel function of the second kind with first order \cite{Gradshteyn_book_2007}, and
\begin{equation}\label{eq:nu_def}
\nu  = {2^{2R}} - 1
\end{equation}
with $R$ denoting the (fixed) transmission rate of the source.
\end{proposition}
\begin{proof}
See Appendix \ref{append:outage_single_relay}. 
\end{proof}

Given the fixed transmission rate $R$ at the source and the outage probability $P_{\rm{out}}^{\rm{HTC}}$, the effective transmission rate can be written as $R\left(1-P_{\rm{out}}^{\rm{HTC}}\right)$. Then, the throughput of the considered system can be obtained by calculating the product of the effective transmission rate and the time of UL information transmission. Based on the approximate expression of the outage probability $P_{\rm{out}}^{\rm{HTC}}$ in (\ref{eq:outage_HTC_AF_prop}), we can have the following corollary on the average throughput of the proposed protocol in the reference model:
\begin{corollary}
An approximate closed-form expression for the average throughput of the HTC protocol in the three-node reference model can be expressed as
\begin{equation}\label{eq:throughput}
{\mathcal T_{\rm{HTC}}} \approx R\left( {1 - P_{{\rm{out}}}^{\rm{HTC}}} \right)\left( {1 - \tau } \right).
\end{equation}
\end{corollary}

Similarly, we can derive the average throughput of the harvest-then-transmit protocol proposed in \cite{Ju_TWC_2014}, where the source transmits its information in the UL without the assistance of the relay. This can be summarized in the following corollary:
\begin{corollary}
The average throughput of the harvest-then-transmit protocol can be expressed as
\begin{equation}\label{eq:throughput_HTT}
{\mathcal T_{\rm{HTT}}} = R\left( {1 - P_{{\rm{out}}}^{\rm{HTT}}} \right)\left( {1 - \tau } \right),
\end{equation}
where $P_{\rm{out}}^{\rm{HTT}} = 1 - {\cal S}\left( {\frac{{4\nu^\prime }}{{\mu^\prime \sigma _{AS}^2\sigma _{SA}^2}}} \right)$ with $\nu^\prime  = {2^{R}} - 1$ and $\mu^\prime = {{\eta  \left({P_A}/{N_0}\right)\tau}}/{{\left( {1 - \tau } \right)}}$.
\end{corollary}

\begin{remark}
It can be observed from (\ref{eq:Pout_def}) that the outage probability and the parameter $\mu$ defined in (\ref{eq:def_mu}) are in inverse proportion since both $\gamma_{SA}$ and $\gamma_{SRA}$ are directly proportional to $\mu$. Moreover, the value of $\mu$ is proportional to that of the time allocation parameter $\tau$. Thus, the larger the value of $\tau$, the lower the outage probability. This observation is understandable since the larger value of $\tau$ means the more harvested energy at the source and relay, which can result in lower outage probability. However, when it comes to the throughput, indefinitely increasing the value of $\tau$ (to 1) may not always be beneficial. Although the larger value of $\tau$ can lead to a higher probability of successful delivery of the source's packet, the effective time that the source can transmit its information (i.e., $1-\tau$) is reduced. \emph{Therefore, there should exist an optimal value of $\tau$ between 0 and 1 such that the throughput of the considered system is highest}. This hypothesis will be validated by numerical results in Section~\ref{sec:num}.

Compared to the harvest-then-transmit protocol, the proposed HTC protocol can potentially improve the network performance, especially for the case where the source-AP link is bad and the relay can help deliver the source information to the AP. But, when the AP-source link is poor and the amount of energy harvested at the source is very small, both the source-AP and source-relay links may suffer from outage. However, since the relay is relatively nearer to the source than the AP, the average outage probability of the source-relay link will be lower than that of the source-AP link. The outage probability of the source-relay link can be further reduced by deploying the relay relatively nearer to the source. In this case, the probability of a good channel condition between the source and relay will be high. Then, the transmission from the source to the relay can still be successful with relatively high probability even when the source only has a limited amount of energy. Thus, we can deduce that the relay should be deployed nearer to the source side to relieve the performance loss due to the cases when the AP-source link is poor. This analysis will also be validated by simulation results in Section~\ref{sec:num}.
$~~\square$
\end{remark}

\section{Extension to Multi-Relay Scenario}
In practice, there may exist more than one relay that are located between the source and AP and would like to assist the communication between them. In this section, we extend our previous analysis to the multi-relay scenario. In conventional multi-relay cooperative networks, the (single) relay selection technique has drawn many efforts since it can improve the network performance while avoiding the reduction in spectral efficiency. Opportunistic relaying\footnote{Note that the OR scheme can achieve the same diversity-multiplexing tradeoff as the space-time coded cooperative diversity scheme, no matter whether the relays implement AF or DF protocols \cite{Bletsas_JSAC_2006}.} (OR) \cite{Bletsas_JSAC_2006} and partial relay selection (PRS) \cite{Krikidis_CL_2008,chen2011exact} are two representative relay selection schemes that were proposed for the AF relaying. In the OR scheme, only the relay that can provide the best end-to-end path between source and destination would be selected from all the available candidates. In contrast, the relay selection procedure of the PRS scheme is performed based on only the channel state information (CSI) of one of the two hops, thereby resulting in lower complexity. Although it has been shown in the literature that the PRS scheme is inferior to the OR scheme, it is still interesting to analyze its performance and compare it with the OR scheme since it has relatively lower complexity \cite{Krikidis_CL_2008}. The throughput performance of the considered WPCCN with these two relay selection protocols will both be analyzed in this section.

We assume that there are $N$ relays located between the source and AP, denoted by $R_i$, $i\in{\mathcal N} = \left\{1,\ldots,N\right\}$. We use ${h_{A R_i }}\sim {\rm {EXP}}\left( {{{\sigma_{ AR_i } ^2}}} \right)$, ${h_{SR_i}}\sim {\rm {EXP}}\left( {{{\sigma_{SR_i} ^2}}} \right)$ and ${h_{R_i A}}\sim {\rm {EXP}}\left( {{{\sigma_{ R_i A} ^2}}} \right)$ to denote the channel power gains from the AP to the relay $R_i$, that from the source to the relay $R_i$, and that from the relay $R_i$ to the AP, respectively. Then, for the relay $R_i$, the instantaneous SNR for the first hop and second hop of the UL information transmission are respectively given by
\begin{equation}\label{eq:SNR_relay_i}
{\gamma _{SR_i}} = \mu {{ {{h_{AS}}} }}{{ {{h_{SR_i}}} }},~{\rm{and}}~
{\gamma _{R_iA}} = \mu {{ {{h_{AR_i}}} }}{{ {{h_{R_iA}}} }}.
\end{equation}

For convenience, we assume that the relays are clustered relatively close together (i.e., location-based clustering). This assumption is commonly used in the context of cooperative diversity systems (e.g., \cite{Krikidis_CL_2008,Costa_SPL_2010} and references therein) and results in the equivalent average channel power gains of the links $A$-$R_i$, $S$-$R_i$ and $R_i$-$A$, respectively. For convenience, we define $\sigma _{A{R_i}}^2 = \sigma _{AR}^2 $, $\sigma _{S{R_i}}^2 = \sigma _{SR}^2$, and $\sigma _{{R_i}A}^2 = \sigma _{RA}^2$ for any $i \in {\mathcal N}$.

To proceed, we first formally describe the selection criteria of the OR and PRS protocols in the following:
\begin{itemize}
  \item In the OR protocol, the relay selection decision is determined by jointly considering the hops $\left\{S\rightarrow R_i\right\}$ and $\left\{R_i \rightarrow A\right\}$. Particularly, the relay with the highest $\min\left( \gamma_{SR_i},\gamma_{R_iA}\right)$ will be selected \cite{Bletsas_JSAC_2006}. We use $R_b$ to denote the selected relay hereafter. Then, the index of the selected relay in the OR protocol is given by
        \begin{equation}\label{eq:OR_criteria}
            b_{\rm{OR}} = \arg {\max}_{i \in {\mathcal N}} \left\{\min \left( {{\gamma _{S{R_i}}},{\gamma _{{R_i}A}}} \right)\right\}.
        \end{equation}
  \item In the PRS scheme, it is assumed that only the CSI of either the hop $\left\{S \rightarrow R_i\right\}$ or the hop $\left\{R_i \rightarrow A\right\}$ is available. The relay selection procedure is modified accordingly such that only one of the two hops is taken into account. Specifically, when only the links in the first hop $\left\{S \rightarrow R_i\right\}$ are considered, the index of the selected relay is determined by \cite{Krikidis_CL_2008}
    \begin{equation}\label{eq:PRS_criteria_1}
        {b_{{\rm{PRS - I}}}} = \arg {\max} _{i \in {\mathcal N}}\left\{\gamma _{S{R_i}}\right\}.
    \end{equation}

    Similarly, for the case that only the second hop $\left\{R_i \rightarrow A\right\}$ is taken into consideration, a single relay is selected according to
    \begin{equation}\label{eq:PRS_criteria_2}
    {b_{{\rm{PRS - II}}}} = \arg {\max} _{i \in {\mathcal N}}\left\{\gamma _{{R_i}A}\right\}.
    \end{equation}
\end{itemize}
\emph{For the purpose of exposition, we consider that the selected relay will only exhaust the energy harvested during the current transmission block to forward the signal received from the source}. In the following two subsections, we will derive the closed-form expressions of the approximate throughput and asymptotic throughput at high SNR for the considered network with OR and PRS protocols, respectively.

\subsection{OR Protocol}
Similar to the analysis in Section \ref{sec:analysis}, we first derive the outage probability of the HTC protocol with OR and obtain the following result:
\begin{proposition}\label{prop:outage_OR}
The approximate outage probability of the OR protocol in the considered WPCCN is given by
\begin{equation}\label{eq:outage_OR_lowerbound}
\begin{split}
&P_{{\rm{out}}}^{{\rm{HTC,OR}}} \approx  1 - {\cal S}\left( {\frac{{4\nu }}{{\mu \sigma _{AS}^2\sigma _{SA}^2}}} \right) +\\
&\sum\limits_{n = 1}^N {N\choose n} {\left( { - 1} \right)^n}{\left[ {{\cal S}\left( {\frac{{4\nu }}{{\mu \sigma _{AR}^2\sigma _{RA}^2}}} \right)} \right]^n} \times \\
&\left[ {{\cal S}\left( {\frac{{4n\nu }}{{\mu \sigma _{AS}^2\sigma _{SR}^2}}} \right) - {\cal S}\left( {\frac{{4n\nu }}{{\mu \sigma _{AS}^2\sigma _{SR}^2}} + \frac{{4\nu }}{{\mu \sigma _{AS}^2\sigma _{SA}^2}}} \right)} \right].
\end{split}
\end{equation}
\end{proposition}
\begin{proof}
See Appendix \ref{append:proof_OR}.
\end{proof}
It can observed that (\ref{eq:outage_OR_lowerbound}) is simplified to (\ref{eq:outage_HTC_AF_prop}) when $N$ is set to 1, which validates our derivation. Based on (\ref{eq:outage_OR_lowerbound}), we can have the following corollary regarding the throughput of the considered WPCCN with OR:
\begin{corollary}
The throughput of the proposed HTC protocol with OR scheme can be approximated as
\begin{equation}\label{eq:throughput_OR}
{\mathcal T}_{\rm{HTC}}^{{\rm{OR}}} \approx R \left(1-P_{{\rm{out}}}^{{\rm{HTC,OR}}}\right) \left(1-\tau\right).
\end{equation}
\end{corollary}

In many practical applications, the system performance at high SNR is of great importance and use. Thus, we investigate the asymptotic throughput of the considered network with OR implemented and have the following proposition:
\begin{proposition}\label{corol:throughput_OR_aymp}
When $P_A/N_0$ is sufficiently large, the throughput in (\ref{eq:throughput_OR}) can be asymptotically expressed as
\begin{equation}\label{eq:throughput_OR_asymp}
{\mathcal T}_{\rm{HTC}}^{{\rm{OR}},asymp} = R \left(1-P_{{\rm{out}}}^{{\rm{OR}},asymp}\right) \left(1-\tau\right),
\end{equation}
where
\begin{equation}\label{}
\begin{split}
&P_{{\rm{out}}}^{{\rm{OR}},asymp} = - {\cal W}\left( {\frac{{4\nu }}{{\mu \sigma _{AS}^2\sigma _{SA}^2}}} \right) + \\
&\sum\limits_{n = 1}^N {N\choose n} {\left( { - 1} \right)^n}{\left[1 + {{\cal W}\left( {\frac{{4\nu }}{{\mu \sigma _{AR}^2\sigma _{RA}^2}}} \right)} \right]^n} \times \\
&\left[ {{\cal W}\left( {\frac{{4n\nu }}{{\mu \sigma _{AS}^2\sigma _{SR}^2}}} \right) - {\cal W}\left( {\frac{{4n\nu }}{{\mu \sigma _{AS}^2\sigma _{SR}^2}} + \frac{{4\nu }}{{\mu \sigma _{AS}^2\sigma _{SA}^2}}} \right)} \right]
\end{split}
\end{equation}
with
\[{\mathcal W}\left( x \right) = \frac{x}{2}\ln \frac{{\sqrt x }}{2}.\]
\end{proposition}
\begin{proof}
See Appendix \ref{proof_corol_throughput_OR_aymp}.
\end{proof}
Note that the asymptotic throughput of the three-node reference model can be obtained by substituting $N=1$ into (\ref{eq:throughput_OR_asymp}).

\subsection{PRS protocol}
To obtain the analytical expression for the throughput of the HTC protocol with PRS schemes, we first investigate the outage probability of these two schemes and have the following proposition:
\begin{proposition}\label{prop:outage_PRS}
The approximate expressions for the outage probability of the PRS protocols with the selection criteria in (\ref{eq:PRS_criteria_1}) and (\ref{eq:PRS_criteria_2}) are respectively given by
\begin{equation}\label{eq:outage_PRS1_lowerbound}
\begin{split}
&P_{{\rm{out}}}^{{\rm{HTC,PRS -I}}} \approx 1 - {\cal S}\left( {\frac{{4\nu }}{{\mu \sigma _{AS}^2\sigma _{SA}^2}}} \right) - \\
&{\cal S}\left( {\frac{{4\nu }}{{\mu \sigma _{AR}^2\sigma _{RA}^2}}} \right)
N\sum\limits_{n = 0}^{N - 1} {{N-1}\choose n}\frac{{{{\left( { - 1} \right)}^n}}}{{n + 1}} \times  \\
&\left[ {{\cal S}\left( {\frac{{4\left( {n + 1} \right)\nu }}{{\mu \sigma _{AS}^2\sigma _{SR}^2}}} \right) - {\cal S}\left( {\frac{{4\left( {n + 1} \right)\nu }}{{\mu \sigma _{AS}^2\sigma _{SR}^2}} + \frac{{4\nu }}{{\mu \sigma _{AS}^2\sigma _{SA}^2}}} \right)} \right],
\end{split}
\end{equation}
\begin{equation}\label{eq:outage_PRS2_lowerbound}
\begin{split}
&P_{{\rm{out}}}^{{\rm{HTC,PRS -II}}} \approx 1 - {\cal S}\left( {\frac{{4\nu }}{{\mu \sigma _{AS}^2\sigma _{SA}^2}}} \right) - \\
&\left\{\sum\limits_{n = 1}^N {{N\choose n}{{\left( { - 1} \right)}^{n + 1}}{{\left[ {{\mathcal S}\left( {\frac{{4\nu }}{{\mu \sigma _{AR}^2\sigma _{RA}^2}}} \right)} \right]}^n}}\right\} \times  \\
&\left[ {{\cal S}\left( {\frac{{4\nu }}{{\mu \sigma _{AS}^2\sigma _{SR}^2}}} \right) - {\cal S}\left( {\frac{{4\nu }}{{\mu \sigma _{AS}^2\sigma _{SR}^2}} + \frac{{4\nu }}{{\mu \sigma _{AS}^2\sigma _{SA}^2}}} \right)} \right].
\end{split}
\end{equation}
\end{proposition}
\begin{proof}
See Appendix \ref{sec:proof_PRS}.
\end{proof}

Subsequently, we can have the following corollary regarding the approximate throughput of the PRS schemes and the corresponding asymptotic performance:
\begin{corollary}
The approximate throughput of the considered system with PRS is given by
\begin{equation}\label{eq:throughput_PRS_ub}
{\mathcal T}_{{\rm{HTC}}}^{{\rm{PRS - X}}} \approx R\left( {1 - P_{{\rm{out}}}^{{\rm{HTC,PRS - X}}}} \right)\left( {1 - \tau } \right),
\end{equation}
where ${\rm{X}} \in \left\{ {{\rm{I,II}}} \right\}$ corresponds to the two different PRS schemes in (\ref{eq:PRS_criteria_1}) and (\ref{eq:PRS_criteria_2}). Moreover, the throughput in (\ref{eq:throughput_PRS_ub}) can be asymptotically expressed as
\begin{equation}\label{eq:throughput_PRS_asymp}
{\mathcal T}_{{\rm{HTC}}}^{{\rm{PRS - X}},asymp} = R\left( {1 - P_{{\rm{out}}}^{{\rm{PRS - X}},asymp}} \right)\left( {1 - \tau } \right),
\end{equation}
where
\begin{equation}\label{eq:outage_PRS1_asymp}
\begin{split}
&P_{{\rm{out}}}^{{\rm{PRS -I,}}asymp} = - {\cal W}\left( {\frac{{4\nu }}{{\mu \sigma _{AS}^2\sigma _{SA}^2}}} \right) - \\
&\left[1+{\cal W}\left( {\frac{{4\nu }}{{\mu \sigma _{AR}^2\sigma _{RA}^2}}} \right)\right]
N\sum\limits_{n = 0}^{N - 1} {{N-1}\choose n}\frac{{{{\left( { - 1} \right)}^n}}}{{n + 1}} \times  \\
&\left[ {{\cal W}\left( {\frac{{4\left( {n + 1} \right)\nu }}{{\mu \sigma _{AS}^2\sigma _{SR}^2}}} \right) - {\cal W}\left( {\frac{{4\left( {n + 1} \right)\nu }}{{\mu \sigma _{AS}^2\sigma _{SR}^2}} + \frac{{4\nu }}{{\mu \sigma _{AS}^2\sigma _{SA}^2}}} \right)} \right],
\end{split}
\end{equation}
\begin{equation}\label{eq:outage_PRS2_asymp}
\begin{split}
&P_{{\rm{out}}}^{{\rm{PRS -II,}}asymp} =  - {\cal W}\left( {\frac{{4\nu }}{{\mu \sigma _{AS}^2\sigma _{SA}^2}}} \right) - \\
&\left\{\sum\limits_{n = 1}^N {{N\choose n}{{\left( { - 1} \right)}^{n + 1}}{{\left[ {1+{\mathcal W}\left( {\frac{{4\nu }}{{\mu \sigma _{AR}^2\sigma _{RA}^2}}} \right)} \right]}^n}}\right\} \times  \\
&\left[ {{\cal W}\left( {\frac{{4\nu }}{{\mu \sigma _{AS}^2\sigma _{SR}^2}}} \right) - {\cal W}\left( {\frac{{4\nu }}{{\mu \sigma _{AS}^2\sigma _{SR}^2}} + \frac{{4\nu }}{{\mu \sigma _{AS}^2\sigma _{SA}^2}}} \right)} \right].
\end{split}
\end{equation}
\end{corollary}

\begin{remark}
In the considered relay selection schemes, only one relay out of the cluster is chosen to assist the source's information transmission in each block. If a relay is not chosen in a certain block, it can store the harvested energy for its own future transmission. Since the channel gains for all relays are assumed to be i.i.d, each relay is selected to help the source with the probability $\frac{1}{N}$. Thus, the average energy harvested by each relay is $\left(1- \frac{1}{N} \right)\eta \tau P_A\sigma_{AR}^2 $. \emph{This amount of energy can actually be regarded as the motivation of the relays to keep active for voluntary cooperation instead of sleeping. On the other hand, the performance of the considered system can be further improved when the energy accumulation at the relays is considered, i.e., the selected relay can use the accumulated energy harvested during the previous non-selected block. In this case, a more sophisticated relay scheduling scheme needs to be designed, which is regarded as our future work and can actually be lower bounded by the schemes investigated in this~paper}.$~~~~~~~~~~~~~~~~~~~~~~~~~~~~~~~~~~~~~~~~~~~~~~~~~~~~~~~~~~\square$
\end{remark}

\section{Numerical Results}\label{sec:num}

In this section, we present some numerical results to illustrate and validate the above theoretical analysis. In the context of wireless energy transfer, distances between nodes are particularly important since they determine not only the reception reliability but also energy attenuation of the received signals. In this paper, even though we do not consider the spatial randomness of node locations \cite{Ding_SPL_2013_Coop,Ding_TWC_2014_Wireless}, we follow some recent papers in the literature (e.g. \cite{Zhou_TWC_2013,Ding_TWC_2014_Power}) and adopt a practical path loss model that captures the effect of node distance on the system performance. This allows us to obtain meaningful insights into the network performance. It is noteworthy that considering the spatial node distributions \cite{Ding_SPL_2013_Coop,Ding_TWC_2014_Wireless} would lead to a more practical framework. However, this will require a new analytical framework, which is out of the scope of this paper and will be treated as our future work.

To capture the effect of node distance on the network performance, we use the channel model that $\sigma _{AS}^2 = \sigma _{SA}^2 = 10^{-3} \left({d_{AS}}\right)^{ - \chi }$, $\sigma _{AR}^2 = \sigma _{RA}^2 = 10^{-3}\left({d_{AR}}\right)^{ - \chi }$, and $\sigma _{SR}^2 = \sigma _{RS}^2 = 10^{-3} \left({d_{SR}}\right)^{ - \chi }$, where $d_{XY}$ denotes the distance between nodes $X$ and $Y$ and $\chi \in [2,5]$ is the path loss factor \cite{Chen_SPL_2010}. Note that a 30dB average signal power attenuation is assumed at a reference distance of 1 meter (m) in the above channel model \cite{Ju_TWC_2014}. A linear topology that the relay(s) is (are) located on a straight line between the source and hybrid AP is considered, i.e, $d_{AR}=  d_{AS} - d_{SR}$. In all simulations, we set the distance between the AP and source $d_{AS} = 10$m, the path loss factor $\chi = 2$, the noise power $N_0 = -80$dBm, the energy harvesting efficiency $\eta = 0.5$, and the fixed transmission rate of the source $R = 1$ bit per channel use (bpcu).

\begin{figure}
\centering \scalebox{0.5}{\includegraphics{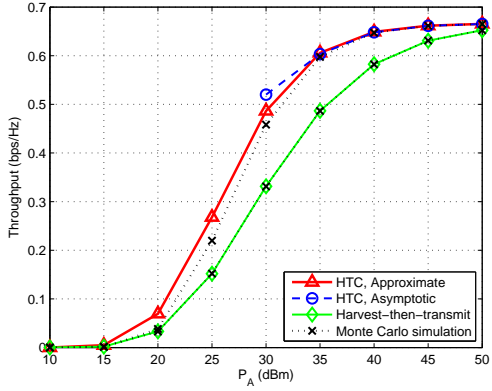}}
\caption{Average throughput of the proposed protocol in the three-node reference model (i.e., $N=1$) and that of harvest-then-transmit protocol versus $P_A$, where $d_{SR}= 3$m, and $\tau = 1/3$. \label{fig:throughput_SNR}}
\end{figure}

\begin{figure}
\centering \scalebox{0.5}{\includegraphics{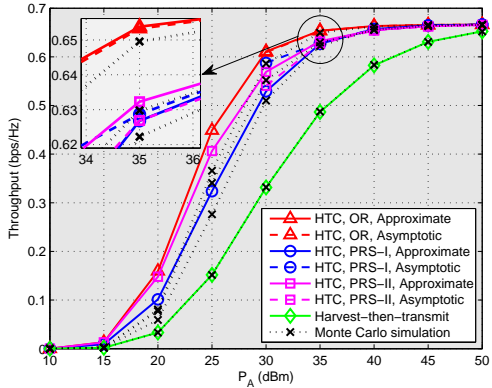}}
\caption{Average throughput of the proposed protocol with different relay selection schemes versus $P_A$, where $N=3$, $d_{SR}= 3$m, and $\tau = 1/3$. \label{fig:throughput_SNR_1}}
\end{figure}

First, we compare the approximate and asymptotic expressions for the average throughput of the considered network derived in the sections above with the corresponding simulation results. Fig. \ref{fig:throughput_SNR} illustrates the throughput performance of the proposed HTC protocol in the three-node reference model and that of the harvest-then-transmit protocol \cite{Ju_TWC_2014} versus $P_A$ for given values of $d_{SR}$, $\tau$ and $R$. We can see from Fig. \ref{fig:throughput_SNR} that the derived approximate expression for the throughput of the proposed protocol becomes very tight in medium and high SNR conditions. Moreover, the asymptotic result\footnote{The asymptotic results in Figs.~\ref{fig:throughput_SNR}-\ref{fig:throughput_SNR_1} were plotted only when the values of $P_A$ exceed certain values to avoid abnormal curves.} tends to coincide with the simulated one at high SNR. Similar phenomenon can also be observed from Fig. \ref{fig:throughput_SNR_1}, in which the throughput curves of the HTC protocol with three different relay selection schemes are plotted for a 3-relay network with the same set of parameters as in Fig. \ref{fig:throughput_SNR}. These observations validate our theoretical analyses. It can also be observed from Figs. \ref{fig:throughput_SNR}-\ref{fig:throughput_SNR_1} that even when there is only one relay node, the proposed HTC protocol can introduce obvious performance gain compared with the harvest-then-transmit protocol. Moreover, this performance gain can be further enlarged in the multi-relay scenario by employing relay selection techniques. From Figs. \ref{fig:throughput_SNR}-\ref{fig:throughput_SNR_1}, we can also observe that the throughput of the HTC protocol tends to be saturated when the SNR is high enough. This observation is understandable since for a given value of $\tau$, the throughput of the system will approach to its maximal value $R\left(1-\tau\right)$ as the outage probability goes to 0 when the SNR is high enough.

\begin{figure}
\centering \scalebox{0.5}{\includegraphics{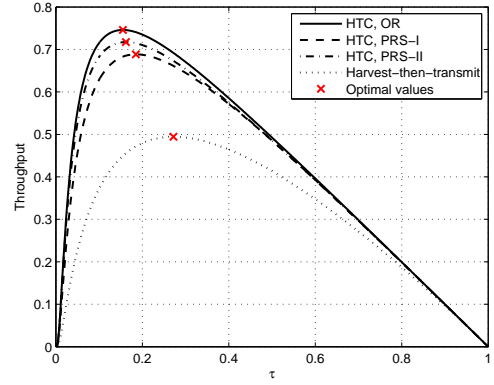}}
\caption{Average throughput versus $\tau$ with $P_A  = 35$dBm, $N = 2$, $d_{SR} = 3$m. \label{fig:throughput_SNR_tau}}
\end{figure}

\begin{figure}
\centering \scalebox{0.5}{\includegraphics{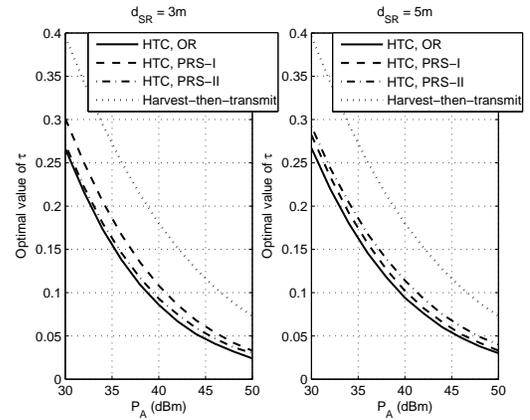}}
\caption{Optimal value of $\tau$ versus $P_A$ with different relay positions, where $N = 2$. \label{fig:optimal_tau_SNR}}
\end{figure}
Next, we will investigate the impacts of the system parameters on the throughput performance in medium and high SNR conditions. Since the theoretical analyses agree well with the simulations in this SNR range, we will only plot the analytical results in the remaining figures. In Fig. \ref{fig:throughput_SNR_tau}, we plot the throughput curves of the HTC protocol and harvest-then-transmit protocol versus the time allocation parameter $\tau$ for a two-relay network with $P_A = 35$dBm and $d_{SR} = 3$m. It can be seen from Fig. \ref{fig:throughput_SNR_tau} that there exists a optimal value of $\tau$ for all schemes. This phenomenon actually coincides with our hypothesis in Remark 1. Moreover, we can observe from Fig. \ref{fig:throughput_SNR_tau} that the optimal values of $\tau$ for the HTC protocol are smaller than that for the harvest-then-transmit protocol. This means that the latter needs to consume more energy at the AP than the former to achieve its maximal throughput. In Fig.~\ref{fig:optimal_tau_SNR}, we evaluate the optimal value of $\tau$ versus the value of $P_A$ for different schemes. Note that the optimal value of $\tau$ can easily be obtained by a one-dimension exhaustive search. It can be observed from Fig. \ref{fig:optimal_tau_SNR} that the optimal values of $\tau$ for all schemes monotonically decrease as $P_A$ increases. This indicates that the higher the SNR, the smaller the optimal value of $\tau$. This observation makes sense since the source can harvest the same amount of energy with shorter time when the AP transmit power increases, and more time could be allocated to the information transmission to improve the throughput. In the considered SNR range, the OR scheme achieves the smallest value of optimal $\tau$, while that of the harvest-then-transmit protocol is the largest one. The optimal $\tau$'s of two PRS schemes lie in the middle, which can be larger than each other depending on the relay position.

\begin{figure}
\centering \scalebox{0.5}{\includegraphics{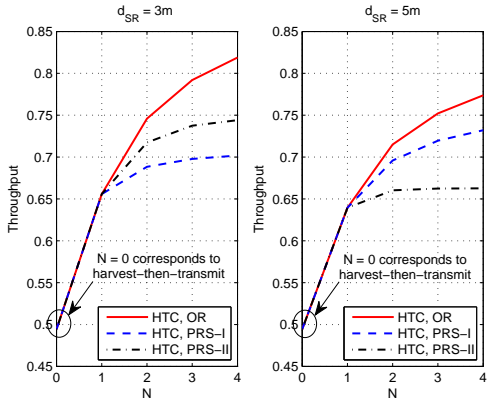}}
\caption{The impact of relay number on the average throughput with optimal $\tau$'s and different relay positions, where $P_A = 35$dBm. \label{fig:throughput_relay_num}}
\end{figure}
\begin{figure}
\centering \scalebox{0.5}{\includegraphics{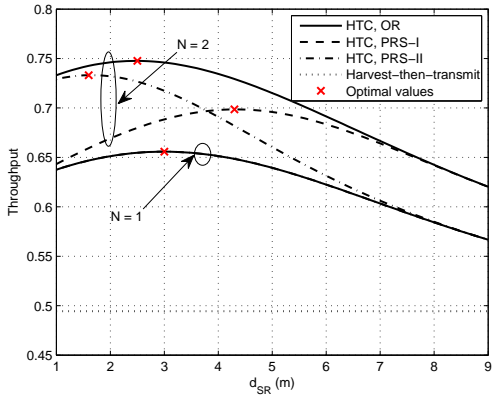}}
\caption{The impact of relay position on the average throughput with optimal $\tau$'s, where $P_A = 35$dBm and $N = 1,~2$. \label{fig:throughput_d_SR}}
\end{figure}

In Fig. \ref{fig:throughput_relay_num}, we demonstrate the influence of the relay number on the throughput performance of three different relay selection schemes with the optimal $\tau$'s and different relay positions. As a comparison, the performance of the harvest-then-transmit protocol is also included in Fig. \ref{fig:throughput_relay_num} as a special case of the HTC protocol for the case that the relay number is zero. We can see from this figure that the curves of the three relay selection schemes coincide when $N =1$. The understanding of this result is straightforward and verifies the correctness of our derivations once again. It can also be observed from Fig. \ref{fig:throughput_relay_num} that the throughput performance of all three schemes is improved when the relay number increases. However, the slope of the performance improvement is decreasing, especially for the PRS schemes. Moreover, it is shown in both subfigures of Fig. \ref{fig:throughput_relay_num} that the OR scheme always achieves the best performance, while the two PRS schemes (i.e., PRS-I and PRS-II) can outperform each other determined by the position of relays. In particular, when the relays are closed to the source (i.e., the case $d_{SR}= 3$m), the PRS-II scheme can achieve better throughput than the PRS-I scheme. However, this phenomenon is reversed when the distance between the source and relays is far enough (e.g., $d_{SR}= 5$m). This observation reveals that in the PRS schemes, the single relay should be selected based on the CSIs of the second hop if the relays are very close to the source, otherwise this procedure should be performed based on the first hop. The reason behind this conclusion is that the performance of the $S$-$R_i$-$A$ paths is actually determined by the weaker~hop.

Fig. \ref{fig:throughput_d_SR} depicts the effect of relay location on the throughput performance of the proposed HTC protocol with different relay selection schemes, in which the throughput curves are plotted versus $d_{SR}$ with the optimal values of $\tau$'s and different relay numbers. As a comparison, we also plot the corresponding throughput of the harvest-then-transmit protocol, which are straight lines since the throughput of the harvest-then-transmit protocol is independent of the relay position. Similar to the observation from Fig. \ref{fig:throughput_relay_num}, we can see from Fig. \ref{fig:throughput_d_SR} that the three HTC schemes coincide when there is only one relay. It is worth mentioning that the optimal value of $d_{SR}$ is around $3$m when $N=1$. This indicates that the single relay should be deployed relatively nearer to the source side, which is in line with our discussions in Remark 1 regarding the relay position. For the case $N=2$, the OR scheme always possesses the best performance, and the performance of two PRS schemes (i.e., PRS-I (\ref{eq:PRS_criteria_1}) and PRS-II (\ref{eq:PRS_criteria_2})) depends on the position of relays. As can also be seen from Fig. \ref{fig:throughput_d_SR}, there exists a throughput-optimal value of $d_{SR}$ for a given relay number. In addition, the optimal value of $d_{SR}$ for the PRS-I scheme is larger than that for the PRS-II scheme. Moreover, with the optimal values of $d_{SR}$, the OR scheme performs the best and the PRS-II scheme performs better than the PRS-I scheme. This observation can be regarded as the tradeoff between the performance and the complexity since these three relay selection schemes require different amounts of CSI. Specifically, the OR scheme needs to collect the most amount of CSI (i.e., $\{h_{AR_i}\}$, $\{h_{SR_i}\}$, and $\{h_{R_iA}\}$). The PRS-II scheme requires a less amount of CSI (i.e., $\{h_{AR_i}\}$ and $\{h_{R_iA}\}$), while the PRS-I scheme need the least amount of CSI (i.e., $\{h_{SR_i}\}$). Finally, jointly considering Figs. \ref{fig:throughput_SNR}-\ref{fig:throughput_d_SR}, it is worth claiming that the proposed HTC protocol is superior to the harvest-then-transmit protocol in all considered cases.

\section{Conclusions}
In this paper, we proposed a harvest-then-cooperate (HTC) protocol for wireless-powered cooperative communication networks. The proposed HTC protocol was first described based on the three-node reference model. We then derived the approximate closed-form expression of the average throughput for the proposed protocol over Rayleigh fading channels. This analysis was subsequently extended to the multi-relay scenario, where the throughput of the proposed protocol with opportunistic relaying and partial relay selection schemes was analyzed. Numerical results showed that the proposed HTC protocol outperforms the harvest-then-transmit one in all simulated scenarios. The performance gain of the proposed HTC protocol can be further improved when the number of relays increases. With the optimal values of the time allocation parameter and relay position, the opportunistic relaying scheme achieves the best throughput performance. Moreover, the partial relay selection scheme based on the second hop can perform better than that based on the first hop.

\begin{appendix}
\subsection{Proof of Proposition \ref{prop:outage_single_relay}}\label{append:outage_single_relay}

To proceed, the mutual information between the source and hybrid AP is given by
\begin{equation}\label{}
{I_{SA}} = \frac{{1 }}{2}{\log _2}\left( {1 + {\gamma _A }} \right).
\end{equation}
Then, we can write the outage probability as
\begin{equation}\label{eq:Pout_def}
\begin{split}
{P_{{\rm{out}}}^{\rm{HTC}} }  &= \Pr \left( {{I_{SA} } < R} \right) = \Pr \left( {{\gamma _A } < \nu} \right)\\
& = \Pr \left( \gamma_{SA} < \nu, \gamma_{SRA}<\nu \right).
\end{split}
\end{equation}
Here, it is worth emphasizing that ${\gamma _{SA}}$ and ${\gamma _{SRA}}$ are correlated since both of them contain the random variable $h_{AS}$. Thus, we have
$\Pr \left( {{\gamma _{SA}} < \nu ,{\gamma _{SRA}} < \nu } \right) \ne \Pr \left( {{\gamma _{SA}} < \nu } \right)\Pr \left( {{\gamma _{SRA}} < \nu } \right)$.
This is in contrast to the conventional constant-powered cooperative networks, where the SNRs of different paths are normally independent. Due to this correlation and the complex structure of ${\gamma _{SRA}}$, it is hard to obtain a closed-form expression of (\ref{eq:Pout_def}). To tackle this, we make the following approximation to the expression of ${\gamma _{SRA}}$ given in (\ref{eq:exact_gamma_SRA}) \cite{Anghel_TWC_2004_Exact,Ikki_CL_2007}:
\begin{equation}\label{eq:approx_gamma_SRA}
\begin{split}
{\gamma _{SRA}} &\approx \frac{{\mu {h_{AS}}{h_{SR}}\mu{h_{AR}}{h_{RA}}}}{{\mu{h_{AS}}{h_{SR}} + \mu{h_{AR}}{h_{RA}}}} \\
&\le \mu \min \left( {{h_{AS}}{h_{SR}},{h_{AR}}{h_{RA}}} \right),
\end{split}
\end{equation}

The approximate SNR value in (\ref{eq:approx_gamma_SRA}) is analytically more tractable than the exact value in (\ref{eq:exact_gamma_SRA}) and thus facilitates the derivation of the closed-form expression of the outage probability. It is also shown that this approximation is accurate enough in medium and high SNR values, which will be verified in Section \ref{sec:num}.

Based on the approximation in (\ref{eq:approx_gamma_SRA}), we can obtain the outage probability of the HTC protocol approximately given~by
\begin{equation}\label{eq:outage_HTC_AF}
\begin{split}
&P_{{\rm{out}}}^{\rm{HTC}} \\
\approx & \Pr \left( {{h_{AS}}{h_{SA}} < \nu /\mu ,\min \left( {{h_{AS}}{h_{SR}},{h_{AR}}{h_{RA}}} \right) < \nu /\mu } \right)\\
=& \Pr \left( {{h_{AS}}{h_{SA}} < \nu /\mu } \right) -\\
&\Pr \left( {{h_{AS}}{h_{SA}} < \nu /\mu ,\min \left( {{h_{AS}}{h_{SR}},{h_{AR}}{h_{RA}}} \right) > \nu /\mu } \right)\\
=& \Pr \left( {{h_{AS}}{h_{SA}} < \nu /\mu } \right) -\Pr \left( {{h_{AR}}{h_{RA}} > \nu /\mu } \right)\times\\
&\Pr \left( {{h_{AS}}{h_{SA}} < \nu /\mu ,{h_{AS}}{h_{SR}} > \nu /\mu } \right),
\end{split}
\end{equation}
where the second equality follows by the fact $\Pr \left( {A,B} \right)$ $=$ $\Pr \left( A \right) - \Pr \left( {A,\bar B} \right)$ \cite[pp. 21]{Papoulis_book_2002} and the last equality holds since $h_{AR}$ and $h_{RA}$ are independent of $h_{AS}$. Next, we calculate the three probabilities in the last step of (\ref{eq:outage_HTC_AF}), respectively. Firstly, we have
\begin{equation}\label{eq:appendix_CDF_1}
\begin{split}
&\Pr \left( {{h_{AS}}{h_{SA}} < \frac{\nu} {\mu}  } \right)= \int_0^\infty  {\Pr \left( {{h_{AS}} < \frac{{\nu }}{\mu y}} \right){f_{{h_{SA}}}}\left( y \right)dy}  \\
=& 1 - \frac{1}{{\sigma _{SA}^2}}\int_0^\infty  {\exp \left( { - \frac{{ \nu }}{{\sigma _{AS}^2 \mu y}} - \frac{y}{{\sigma _{SA}^2}}} \right)} dy \\
=& 1 - {\cal S}\left( {\frac{{4\nu }}{{\mu \sigma _{AS}^2\sigma _{SA}^2}}} \right) ,
\end{split}
\end{equation}
where \cite[Eq. (3.324.1)]{Gradshteyn_book_2007} is used to solve the second integral. Similarly, we have
\begin{equation}\label{eq:appendix_CDF_2}
\Pr \left( {{h_{AR}}{h_{RA}} > \nu /\mu } \right) = {\cal S}\left( {\frac{{4\nu }}{{\mu \sigma _{AR}^2\sigma _{RA}^2}}} \right).
\end{equation}
For the last term, we have
\begin{equation}\label{eq:appendix_CDF_3}
\begin{split}
&\Pr \left( {{h_{AS}}{h_{SA}} <\nu/ \mu  ,{h_{AS}}{h_{SR}} >\nu/ \mu } \right) \\
%=& \int_0^\infty  {\Pr \left( {{h_{SA}} < \frac{{\nu  }}{\mu y},{h_{SR}} > \frac{{\nu }}{\mu y}} \right)} {f_{{h_{AS}}}}\left( y \right)dy \\
=& \int_0^\infty  {\Pr \left( {{h_{SA}} < \frac{{ \nu }}{\mu y}} \right)\Pr \left( {{h_{SR}} > \frac{{\nu }}{\mu y}} \right)} {f_{{h_{AS}}}}\left( y \right)dy \\
=& \int_0^\infty  {\left( {1 - \exp \left( { - \frac{{\mu \gamma }}{{\sigma _{SA}^2y}}} \right)} \right)\exp \left( { - \frac{{\mu \gamma }}{{\sigma _{SR}^2y}}} \right)} {f_{{h_{AS}}}}\left( y \right)dy \\
=& \frac{1}{{\sigma _{AS}^2}}\int_0^\infty  {\exp \left( { - \frac{{\nu }}{{\sigma _{SR}^2 \mu y}} - \frac{y}{{\sigma _{AS}^2}}} \right)} dy -\\
& \frac{1}{{\sigma _{AS}^2}}\int_0^\infty  {\exp \left( { - \frac{1}{y}\left( {\frac{{\nu }}{{\sigma _{SA}^2}\mu} + \frac{{\nu }}{{\sigma _{SR}^2}\mu}} \right) - \frac{y}{{\sigma _{AS}^2}}} \right)} dy \\
=&{{\cal S}\left( {\frac{{4\nu }}{{\mu \sigma _{AS}^2\sigma _{SR}^2}}} \right) - {\cal S}\left( {\frac{{4\nu }}{{\mu \sigma _{AS}^2\sigma _{SR}^2}} + \frac{{4\nu }}{{\mu \sigma _{AS}^2\sigma _{SA}^2}}} \right)} .
\end{split}
\end{equation}

By substituting (\ref{eq:appendix_CDF_1})-(\ref{eq:appendix_CDF_3}) into (\ref{eq:outage_HTC_AF}), we obtain the desired result in (\ref{eq:outage_HTC_AF_prop}). This completes the proof.

\subsection{Proof of Proposition \ref{prop:outage_OR}}\label{append:proof_OR}

According to the selection criterion of the OR protocol given in (\ref{eq:OR_criteria}) and the SNR approximation in (\ref{eq:approx_gamma_SRA}), the outage probability of the OR protocol can be approximately given by (\ref{eq:outage_OR_def}) on top of next page.
\begin{figure*}
\begin{equation}\label{eq:outage_OR_def}
\begin{split}
 P_{{\rm{out}}}^{{\rm{HTC,OR}}} & \approx \Pr \left( {{h_{AS}}{h_{SA}} < \frac{\nu }{\mu },\min \left( {{h_{AS}}{h_{S{R_b}}},{h_{A{R_b}}}{h_{{R_b}A}}} \right) < \frac{\nu }{\mu }} \right) \\
&= \Pr \left( {{h_{AS}}{h_{SA}} < \frac{\nu }{\mu },{{\max }_{i \in {\mathcal N}}}\left\{ {\min \left( {{h_{AS}}{h_{S{R_i}}},{h_{A{R_i}}}{h_{{R_i}A}}} \right)} \right\} < \frac{\nu }{\mu }} \right) \\
&= \int_0^\infty  {\underbrace {\Pr \left( {y{h_{SA}} < \frac{\nu }{\mu }} \right)}_{{{\mathcal T}_1}}\underbrace {\Pr \left( {{{\max }_{i \in {\mathcal N}}}\left\{ {\min \left( {y{h_{S{R_i}}},{h_{A{R_i}}}{h_{{R_i}A}}} \right)} \right\} < \frac{\nu }{\mu }} \right)}_{{{\mathcal T}_2}}} {f_{{h_{AS}}}}\left( y \right)dy. \\
\end{split}
\end{equation}
\hrulefill
\vspace*{4pt}
\end{figure*}
To proceed, it is straightforward to obtain
\begin{equation}\label{eq:expression_T1}
{{\mathcal T}_1} = \Pr \left( {{h_{SA}} < \frac{\nu }{\mu }\frac{1}{y}} \right) = 1 - \exp \left( { - \frac{\nu }{{\mu \sigma _{SA}^2}}\frac{1}{y}} \right).
\end{equation}
For the term ${\mathcal T}_2$ in (\ref{eq:outage_OR_def}), we have
\begin{equation}
\begin{split}
{{\mathcal T}_2} &= \Pr \left( {\min \left( {y{h_{S{R_1}}},{h_{A{R_1}}}{h_{{R_1}A}}} \right) < \frac{\nu }{\mu }} \right) \times  \ldots  \times \\
&~~~~\Pr \left( {\min \left( {y{h_{S{R_N}}},{h_{A{R_N}}}{h_{{R_N}A}}} \right) < \frac{\nu }{\mu }} \right) \\
&= {\left[ {\Pr \left( {\min \left( {y{h_{S{R_1}}},{h_{A{R_1}}}{h_{{R_1}A}}} \right) < \frac{\nu }{\mu }} \right)} \right]^N} \\
&= {\left[ {1 - \Pr \left( {\min \left( {y{h_{S{R_1}}},{h_{A{R_1}}}{h_{{R_1}A}}} \right) > \frac{\nu }{\mu }} \right)} \right]^N}, 
\end{split}
\end{equation}
where the second equation follows since the random variables $\left\{ {{h_{S{R_i}}}} \right\}$, $\left\{ {{h_{A{R_i}}}} \right\}$, and $\left\{ {{h_{{R_i}A}}} \right\}$ are respectively independent and identically distributed (i.i.d). Since the variables ${h_{S{R_1}}}$ and ${h_{A{R_1}}}{h_{{R_1}A}}$ are independent, we further have
\begin{equation}\label{eq:expression_T2}
{{\mathcal T}_2}= {\left[ {1 - \exp \left( { - \frac{\nu }{{\mu \sigma _{SR}^2}}}{\frac{1}{y}}
 \right)\underbrace {\Pr \left( {{h_{A{R_1}}}{h_{{R_1}A}} > \frac{\nu }{\mu }} \right)}_{{\mathcal T}_3}} \right]^N},
\end{equation}
Note that the term ${\mathcal T}_3$ can be written as
\begin{equation}\label{eq:expression_T3}
\begin{split}
{\mathcal T}_3
= {\mathcal S}\left( {\sqrt {\frac{{4\nu }}{{\mu \sigma _{AR}^2\sigma _{RA}^2}}} } \right).
\end{split}
\end{equation}
Substituting (\ref{eq:expression_T3}) into (\ref{eq:expression_T2}) and expanding the term ${\mathcal T}_2$ based on the binomial theorem \cite[Eq. (3.1.1)]{Abramowitz_book_1972}, we can further write (\ref{eq:expression_T2}) as
\begin{equation}\label{eq:expression_T2_expand}
\begin{split}
 {{\cal T}_2} =& {\left[ {1 - \exp \left( { - \frac{\nu }{{\mu \sigma _{SR}^2}}\frac{1}{y}} \right){\cal S}\left( {\frac{{4\nu }}{{\mu \sigma _{AR}^2\sigma _{RA}^2}}} \right)} \right]^N} \\
= &1{\rm{ + }}\sum\limits_{n = 1}^N {N \choose n} {\left( { - 1} \right)^n}\times\\
&{\left[ {{\cal S}\left( {\frac{{4\nu }}{{\mu \sigma _{AR}^2\sigma _{RA}^2}}} \right)} \right]^n}\exp \left( { - \frac{{n\nu }}{{\mu \sigma _{SR}^2}}\frac{1}{y}} \right).
\end{split}
\end{equation}

Substituting (\ref{eq:expression_T1}) and (\ref{eq:expression_T2_expand}) into (\ref{eq:outage_OR_def}), we have (\ref{eq:outage_OR_expression}) on top of next page.
\begin{figure*}
\begin{equation}\label{eq:outage_OR_expression}
\begin{split}
 P_{{\rm{out}}}^{{\rm{HTC,OR}}}  \approx &\int_0^\infty  {\frac{1}{{\sigma _{AS}^2}}\exp \left( { - \frac{y}{{\sigma _{AS}^2}}} \right)dy}  - \frac{1}{{\sigma _{AS}^2}}\int_0^\infty  {\exp \left( { - \frac{y}{{\sigma _{AS}^2}} - \frac{\nu }{{\mu \sigma _{SA}^2}}\frac{1}{y}} \right)} dy + \\
 & \sum\limits_{n = 1}^N {N\choose n} {\left( { - 1} \right)^n}{\left[ {{\cal S}\left( {\frac{{4\nu }}{{\mu \sigma _{AR}^2\sigma _{RA}^2}}} \right)} \right]^n}\frac{1}{{\sigma _{AS}^2}}\left\{ {\int_0^\infty  {\exp \left( { - \frac{y}{{\sigma _{AS}^2}} - \frac{{n\nu }}{{\mu \sigma _{SR}^2}}\frac{1}{y}} \right)} dy - } \right. \\
&\left. {\int_0^\infty  {\exp \left( { - \frac{y}{{\sigma _{AS}^2}} - \frac{{n\nu }}{{\mu \sigma _{SR}^2}}\frac{1}{y} - \frac{\nu }{{\mu \sigma _{SA}^2}}\frac{1}{y}} \right)} dy} \right\}.
\end{split}
\end{equation}
\hrulefill
\vspace*{4pt}
\end{figure*}
Solving the integrals in (\ref{eq:outage_OR_expression}) by following the routine in the proof of Proposition \ref{prop:outage_single_relay}, we obtain the desired result in (\ref{eq:outage_OR_lowerbound}).

\subsection{Proof of Proposition \ref{corol:throughput_OR_aymp}}\label{proof_corol_throughput_OR_aymp}

To obtain the asymptotic analysis of the throughput, we need to perform the asymptotic analysis for the function ${\mathcal S}(x)$ defined in (\ref{eq:S_func}). Moreover, it is straightforward to observe that all terms inside the function ${\mathcal S}(\cdot)$ in (\ref{eq:outage_OR_lowerbound}) approach to zero when $P_A/N_0 \rightarrow \infty$. Thus, to obtain the asymptotic performance of the throughput, we should find the approximation of ${\mathcal S}(x)$ for $x \rightarrow 0$.

By applying the series representation of the modified Bessel function \cite[Eq. (8.446)]{Gradshteyn_book_2007}, we can re-write the function ${\mathcal S}(x)$ as follows:
\begin{equation}
\begin{split}
{\mathcal S}\left( x \right) =& 1 + \sqrt x {I_1}\left( {\sqrt x } \right)\left( {\ln \frac{{\sqrt x }}{2} + \pmb C} \right) -\\
&\frac{1}{2}\sum\limits_{l = 0}^\infty  {\frac{{{{\left( {\frac{{\sqrt x }}{2}} \right)}^{2l + 1}}\sqrt x }}{{l!\left( {l + 1} \right)!}}\left( {\sum\limits_{k = 1}^l {\frac{1}{k}}  + \sum\limits_{k = 1}^{l + 1} {\frac{1}{k}} } \right)}, 
\end{split}
\end{equation}
where $I_1(\cdot)$ is the modified Bessel function of the first kind with first order and $\pmb C$ is the Euler's constant \cite[Eq. (8.367)]{Gradshteyn_book_2007}. Based on the fact that $I_1(x) \rightarrow \frac{x}{2}$ if $x \rightarrow 0$ \cite[Eq. (8.447)]{Gradshteyn_book_2007}, we can obtain the following approximation
\begin{equation}\label{eq:kappa_func_approx}
\begin{split}
{\mathcal S}\left( x \right) \approx& 1 + \frac{x}{2}\left( {\ln \frac{{\sqrt x }}{2} + \pmb C - \frac{1}{2}} \right) - \\
&\frac{1}{2}\sum\limits_{l = 1}^\infty  {\frac{{{{\left( {\frac{{\sqrt x }}{2}} \right)}^{2l + 1}}\sqrt x }}{{l!\left( {l + 1} \right)!}}\left( {\sum\limits_{k = 1}^l {\frac{1}{k}}  + \sum\limits_{k = 1}^{l + 1} {\frac{1}{k}} } \right)}\\
\approx& 1 + \frac{x}{2} {\ln \frac{{\sqrt x }}{2} } ,
\end{split}
\end{equation}
where the second approximation follows since the term $\pmb C - \frac{1}{2}$ is negligible compared with the term $\ln \frac{{\sqrt x }}{2}$ for $x \rightarrow 0$ and the terms in the summation are with higher orders.

\begin{figure}
\centering \scalebox{0.5}{\includegraphics{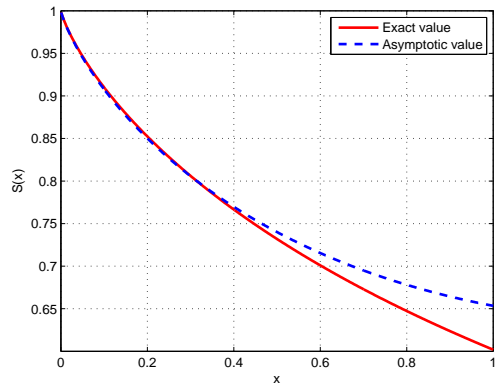}}
\caption{Verification of the approximation for the function ${\mathcal S}\left(x\right)$. \label{fig:approx_bessel_function}  }
\end{figure}

In Fig. \ref{fig:approx_bessel_function}, we plot the exact and asymptotic values of the function ${\mathcal S}(x)$ based on the formulas in (\ref{eq:S_func}) and (\ref{eq:kappa_func_approx}). As can be observed in Fig. \ref{fig:approx_bessel_function}, the exact and asymptotic values coincide with each other very well when $x$ approaches zero, which validates the approximation in (\ref{eq:kappa_func_approx}).

By using (\ref{eq:kappa_func_approx}) and performing some simplifications, we can write the asymptotic throughput of the OR scheme at high SNR as (\ref{eq:throughput_OR_asymp}).

\subsection{Proof of Proposition \ref{prop:outage_PRS}}\label{sec:proof_PRS}
Analogous to the proof of Proposition \ref{prop:outage_single_relay}, we can obtain an approximate expression for the outage probability of the PRS protocols given by
\begin{equation}\label{eq:outage_PRS1_def}
\begin{split}
&P_{{\rm{out}}}^{{\rm{HTC,PRS}}} \approx \Pr \left( {{h_{AS}}{h_{SA}} < \frac{\nu }{\mu }} \right) - \Pr \left( {{h_{A{R_b}}}{h_{{R_b}A}} > \frac{\nu }{\mu }} \right) \times\\
&{\Pr \left( {{h_{AS}}{h_{SA}} < \frac{\nu }{\mu },{h_{AS}}{h_{S{R_b}}} > \frac{\nu }{\mu }} \right)}.
\end{split}
\end{equation}

For the case that the relay is selected only based on the hop $\left\{S\rightarrow R_i\right\}$ with the selection criterion given in (\ref{eq:PRS_criteria_1}), we can claim that the selected relay $R_b$ should meet the following condition
\begin{equation}\label{}
{h_{S{R_b}}} = {\max} _{i \in {\mathcal N}}\left\{ {{h_{S{R_i}}}} \right\}.
\end{equation}
Since the variables $\left\{ {{h_{S{R_i}}}} \right\}$'s are i.i.d, we can obtain the PDF of ${h_{S{R_b}}}$ given by
\begin{equation}\label{eq:PDF_h_SRb}
{f_{{h_{S{R_b}}}}}\left( z \right){ = }\frac{N}{{\sigma _{SR}^2}}\sum\limits_{n = 0}^{N - 1} {{{N-1}\choose n} {{\left( { - 1} \right)}^n}\exp \left( { - \frac{{n + 1}}{{\sigma _{SR}^2}}z} \right)} .
\end{equation}
Moreover, the random variables $h_{AR_b}$ and $h_{R_bA}$ follow the exponential distribution with average values ${\sigma _{AR}^2}$ and ${\sigma _{RA}^2}$, since the links $A$-$R_i$ and $R_i$-$A$ are not taken into account for relay selection. Substituting the PDFs into (\ref{eq:outage_PRS1_def}) and solving the integrals, we can obtain the desired result in (\ref{eq:outage_PRS1_lowerbound}).

When it comes to the PRS protocol with the criterion in (\ref{eq:PRS_criteria_2}), the selected relay $R_b$ should satisfy
\begin{equation}\label{}
{h_{A{R_b}}}{h_{{R_b}A}} = {\max} _{i \in {\mathcal N}}\left\{ {{h_{A{R_i}}}{h_{{R_i}A}}} \right\}.
\end{equation}
Thus, we have
\begin{equation}\label{}
\begin{split}
&\Pr \left( {{h_{A{R_b}}}{h_{{R_b}A}} > \frac{\nu }{\mu }} \right)= 1- \Pr \left( {{h_{A{R_b}}}{h_{{R_b}A}} < \frac{\nu }{\mu }} \right)\\
&=1-{\left[ 1- {{\cal S}\left( {\frac{{4\nu }}{{\mu \sigma _{AR}^2\sigma _{RA}^2}}} \right)} \right]^N}\\
&=\sum\limits_{n = 1}^N {{N\choose n}{{\left( { - 1} \right)}^{n + 1}}{{\left[ {{\mathcal S}\left( {\frac{{4\nu }}{{\mu \sigma _{AR}^2\sigma _{RA}^2}}} \right)} \right]}^n}}.
\end{split}
\end{equation}
In contrast, the random variable $h_{SR_b}$ only follows an exponential distribution with average values ${\sigma _{SR}^2}$. This is because the links $S$-$R_i$'s are not taken into consideration during the relay selection process. Then, the expression (\ref{eq:outage_PRS2_lowerbound}) can be obtained by repeating the integral calculation in (\ref{eq:outage_PRS1_def}). This completes the proof.

\end{appendix}

\ifCLASSOPTIONcaptionsoff
  \newpage
\fi

\section{Acknowledgement}
The authors would like to thank the anonymous reviewers for their valuable comments and suggestions, which improved the quality of the paper. The authors also thank Dr. Jun Li and Yifan Gu for their helpful discussion.

\bibliographystyle{IEEEtran}
\bibliography{References}

\end{document}